\documentclass[12pt, a4paper]{article}
\usepackage{hyperref}
\usepackage{epstopdf}
\usepackage{stackrel}
\usepackage{graphicx}
\usepackage[utf8]{inputenc}
\usepackage{amssymb}
\usepackage{amsthm}
\usepackage{amsmath, calc}
\usepackage[authoryear]{natbib}

\usepackage{xcolor}
\usepackage{subcaption}
\usepackage[boxruled,linesnumbered]{algorithm2e}
\usepackage{adjustbox}
\usepackage{booktabs}
\usepackage{multirow}
\usepackage{rotating}

\usepackage[a4paper,left=3cm,right=2cm,top=2.5cm,bottom=2.5cm]{geometry}

\makeatletter
\newcommand*{\centerfloat}{%
  \parindent \z@
  \leftskip \z@ \@plus 1fil \@minus \textwidth
  \rightskip\leftskip
  \parfillskip \z@skip}
\makeatother

\usepackage[figurename=Figure]{caption}
\usepackage[tablename=Table]{caption}

 \usepackage{color}
\usepackage{xcolor}

\begin{document}

\author{
Titin Agustin Nengsih\\
IRMA, UMR 7501, Universit\'e de Strasbourg, \\
67084 Strasbourg, Cedex, France; \\
and iCUBE, UMR 7357, \\67400 Strasbourg, France. \\
E-mail: nengsih@unistra.fr\\
\\
Fr\'ed\'eric Bertrand\\
IRMA, UMR 7501,\\ Universit\'e de Strasbourg,\\ 
67084 Strasbourg, Cedex, \\
France. E-mail: fbertran@unistra.fr\\
\\
Myriam Maumy-Bertrand\\
IRMA, UMR 7501,\\ Universit\'e de Strasbourg,\\ 
67084 Strasbourg, Cedex, \\
France. E-mail: mmaumy@unistra.fr\\
\\
Nicolas Meyer\\
GMRC, Public Health Department, \\
Strasbourg University Hospital, \\
Strasbourg, France; \\
and iCUBE, UMR 7357, \\
67400 Strasbourg, \\
France. E-mail: meyer@unistra.fr
}

  \title{Determining the Number of Components in PLS Regression on Incomplete Data}

\maketitle

  \abstract{Partial least squares regression---or PLS---is a multivariate method in which models are estimated using either the SIMPLS or NIPALS algorithm. PLS regression has been extensively used in applied research because of its effectiveness in analysing relationships between an outcome and one or several components. Note that the NIPALS algorithm is able to provide estimates on incomplete data. Selection of the number of components used to build a representative model in PLS regression is an important problem. However, how to deal with missing data when using PLS regression remains a matter of debate. Several approaches have been proposed in the literature, including the $Q^2$ criterion, and the $AIC$ and $BIC$ criteria. Here we study the behavior of the NIPALS algorithm when used to fit a PLS regression for various proportions of missing data and for different types of missingness. We compare criteria for selecting the number of components for a PLS regression on incomplete data and on imputed datasets using three imputation methods: multiple imputation by chained equations, $k$-nearest neighbor imputation, and  singular value decomposition imputation. Various criteria were tested with different proportions of missing data (ranging from 5\% to 50\%) under different missingness assumptions. $Q^2$-leave-one-out component selection methods gave more reliable results than $AIC$ and $BIC$-based ones.}

\section{Introduction}
\label{intro}
Missing data are present in many real-world datasets and often cause  problems in data analysis.
Missing data can occur for many reasons, including uncollected data, mishandled samples, equipment errors, measurement errors, misunderstanding questionnaires, etc.\mbox{\citep{Grung1998,Folch-Fortuny2016}.} According to \cite{Little2002}, missing data can be divided into three categories, namely: missing completely at random (MCAR), missing at random (MAR), and missing not at random (MNAR). If the probability that the data is known depends neither on the observed value, nor on the missing values, the data are said to be MCAR. In the case of MAR, missingness depends only on the values of the observed data. Lastly, data are said to be MNAR if missingness depends both on the observed and missing data values.

Many methods have been proposed for imputing missing data. The simplest ones rely on single value imputation, e.g., the mean over the complete cases in the study sample---known as mean imputation \citep{Troyanskaya2001}. More complex methods include regression-based imputation \citep{Horton2001}, imputation based on non-linear iterative partial least squares (NIPALS) \citep{Tenenhaus1998}, multiple imputation \citep{Rubin1987}, $k$-nearest neighbors imputation (KNNimpute) \citep{Dixon1979}, singular value decomposition-based imputation (SVDimpute) \citep{Troyanskaya2001}, and so on.

Partial least squares regression (PLS) was introduced in the 1970s by \cite{Wold1966}. It has gone from being popular in chemometrics \citep[see][]{Wold2001a} to being commonly used in many research areas such as bioinformatics \citep{Nguyen2004a}, medicine \citep{Yang2017}, social sciences \citep{Sawatsky2015}, and spectroscopy \citep{Oleszko2017}. PLS regression---in its classical form---is based on the NIPALS algorithm. The alternative estimation method for PLS regression is SIMPLS algorithm, for \emph{straightforward implementation of a statistically inspired modification to PLS} \citep[see][]{DeJong1993a}. The former has been implemented in software such as SIMCA \citep{L2001} and more recently in the \texttt{plsRglm} R package \citep{Bertrand2015}.

The NIPALS algorithm was initially devised to carry out principal component analysis (PCA) on incomplete datasets. This explains why its reliability under increasing proportions of missing data has been studied mainly in this setting \mbox{\citep{Nelson1996,Grung1998,Arteaga2002}.} 

As in PCA, one of the justifications for using the NIPALS algorithm in PLS regression is that it enables models to be fitted on incomplete datasets. This is long-known and frequently used as an argument to preferentially apply this algorithm. In this paper, we focus on univariate PLS, also known as PLS1, and the use of NIPALS.

Determining the optimal number of components is a very important problem in PLS regression. Selecting a less-than-optimal number of components leads to a loss of information, whereas selecting a more-than-optimal  number can lead to models with poor predictive ability \citep{Wiklund2007}. Several papers have studied ways to determine the number of components to retain in the final PLS regression \citep[see for instance][]{Lazraq2003}.

Though it is now considered a benchmark for incomplete dataset analysis, the reliability of the NIPALS algorithm when estimating PLS regression parameters on incomplete data sets has been studied very little, despite its importance. In the context of PLS regression, details pertaining to missing data, such as how to estimate scores on incomplete data, and the impact of missing data on PLS prediction have been reported by \citet{Nelson1996}, \citet{Rannar1995}, and \citet{Serneels2008}. However, the sensitivity of the NIPALS algorithm to increasing missing data proportions does not seem to have been given much attention. Moreover, in the few papers that pertain to incomplete data issues, the reliability of the NIPALS algorithm under different missingness mechanisms described in \citet{Little1987} has been systematically ignored.

In summary, at least two things may affect parameter estimates for PLS regression: the proportion of missing data, and the type of missingness. Both issues will be studied here. 

In addition, we compare criteria for selecting the number of components in PLS regression. The most-used method for incomplete data sets is PLS regression with the NIPALS algorithm (NIPALS-PLSR). Other methods for imputing data sets are: multiple imputation by chained equations (MICE), $k$-nearest neighbors imputation (KNNimpute), and singular value decomposition imputation (SVDimpute). The influence of the proportion of missing data and of the type of missingness on the estimation of the number of components in a PLS regression is the main purpose of the present study.

The paper is organized as follows. Section~\ref{sec:1} describes the methods, presenting a brief description of PLS regression, cross-validation with missing values, and imputation methods. Section~\ref{sec:2} describes the simulation study, and Section~\ref{sec:3} gives the results of this study. We conclude with a general discussion in Section~\ref{sec:4}.

\section{Partial least squares regression and related works}
\label{sec:1}

\subsection{PLSR-NIPALS}

A complete description of PLS regression can be found in the articles \citet{Wold1966} and \citet{Hoskuldsson1988}. These are summarized here.

\begin{sloppypar}
	Suppose that $\mathbf{X}$ is an $n \times p$ data matrix of continuous explanatory variables $\textbf{x}_1,\dots,\textbf{x}_j,$ $\dots,\textbf{x}_p$ where $p$ can be greater than $n$ (the number of observations). Moreover, the rank of the matrix is equal to $H$. The $p$ variables are supposed centered and scaled to unit variance. This matrix is used to predict a continuous response variable $\mathbf{y}$ with dimensions $n \times 1$. If $p>n$, then a multiple linear regression can not be fitted because the $p \times p$  covariance matrix is singular.
\end{sloppypar}

PLS regression yields orthogonal components, denoted $\textbf{t}_h$, obtained by maximizing the covariance between linear combinations of the columns $\textbf{X}_h$ and the response variable $\mathbf{y}$. Note that $\mathbf{y}$ is also supposed centered and scaled to unit variance. The PLS components are linear combinations of the $p$ variables $\textbf{x}_1,\dots,\textbf{x}_j,\dots,\textbf{x}_p$ that maximize their covariance with $\mathbf{y}$. Let $\textbf{T}$ be the matrix formed by these components $\textbf{t}_1,\ldots,\textbf{t}_h$. The model is then:
\begin{equation}
	\mathbf{y}=\textbf{T}^t \mathbf{c}+\mathbf{\epsilon},
\end{equation}
where the vector $\mathbf{\epsilon}$ represents the residual vector, the vector $\mathbf{c}$ is formed by the regression coefficients of the components $\textbf{t}_h$, and the exponent $t$ means the transpose.

In \citet{Tenenhaus1998}, the NIPALS algorithm divides the dataset into a complete part and an incomplete part to build a  matrix, called $\textbf{X}_h$. The columns of $\textbf{X}_h$ are noted $\textbf{x}_{h1},\ldots,\textbf{x}_{hj},$ $\ldots,\textbf{x}_{hp}$.

The NIPALS algorithm steps on the complete data set are:
\begin{enumerate}    
	\item  Initialize $\textbf{X}_0=\textbf{X}$
	\item For $h=1, 2,\ldots, H$:
	\begin{enumerate}
		\item  $\textbf{t}_h$ = the first column of $\textbf{X}_{h-1}$
		\item  Repeat until convergence:
		\begin{enumerate}
			\item  $\textbf{q}_h = \textbf{X}_{h-1}^t\textbf{t}_h / \textbf{t}_h^t \textbf{t}_h$
			\item  Normalize $\textbf{q}_h$ to 1
			\item  $\textbf{t}_h = \textbf{X}_{h-1}^t\textbf{q}_h / \textbf{q}_h^t \textbf{q}_h$
		\end{enumerate}
			\item   $\textbf{X}_h = \textbf{X}_{h-1} - \textbf{t}_h \textbf{q}_h^t $
	\end{enumerate}
\end{enumerate}		

However, the interest of the NIPALS algorithm in the PLS setting is clearer in the presence of incomplete data. The steps of the  NIPALS algorithm on an incomplete dataset are as follows:
\begin{enumerate}    
	\item  Initialize $\textbf{X}_0=\textbf{X}$
	\item For $h=1,\ldots, H$:
	\begin{enumerate}
		\item  $\textbf{t}_h$ = the first column of $\textbf{X}_{h-1}$
		\item  Repeat until convergence:
		\begin{enumerate}
			\item  For $j=1,\ldots, p$:\\
			$$q_{hj}=\frac{\sum_{i\in I}{ {x}_{h-1,ji}{t}_{hi}} }{ \sum_{i\in I}{{t}_{hi}^{2}}},  \quad \textup{with} \quad I=\left\{i:{x}_{ji} \textup{ and } {t}_{hi} \textup{ are not missing} \right\} $$
			\item  Normalize $\textbf{q}_h$ to 1
			\item  For $i=1,\ldots, n$:
			$$t_{hi}=\frac{\sum_{j\in J}{ {x}_{h-1,ji}{q}_{hj}} }{\sum_{j\in J} {{q}_{hj}^{2}}},\quad \textup{with} \quad J=\left\{j:{x}_{ji} \textup{ is not missing}\right\} $$
		\end{enumerate}
		\item 	$\textbf{X}_{h} = \textbf{X}_{h-1}-\textbf{t}_h\textbf{q}_h^t$ 
	\end{enumerate}
\end{enumerate}	

The NIPALS algorithm also makes it possible to estimate the missing data using the reconstruction formula up to order $h$:
\begin{equation}
	{\hat{x}}_{ij}=\sum_{l=1}^{h}{{t}_{li}{q}_{lj}}.
\end{equation}

NIPALS-PLSR starts with (optionally) transformed, scaled, and centered matrices $\mathbf{X}$ and $\mathbf{y}$, and proceeds using Tenenhaus's algorithm in \cite{Tenenhaus1998}. The component $\textbf{t}_h$ is equal to $\textbf{X}\textbf{w}_h$,  where the weight $\textbf{w}_h$ is constructed step by step. The PLS components are linear combinations of the $p$ variables of the matrix $\textbf{X}$ that maximize the covariance between $\textbf{Xw}$ and $\textbf{y}$. The steps are as follows:
\begin{enumerate}    
	\item Initialize $\textbf{X}_0=\textbf{X}$, $\textbf{y}_0=\textbf{y}$
	\item Repeat until convergence of $\textbf{q}_h$, for $h=1, 2,\ldots, H$:
	\begin{enumerate}
		\item  $\textbf{w}_h = \textbf{X}_{h-1}^t\textbf{y}_{h-1} / \textbf{y}_{h-1}^t\textbf{y}_{h-1}$
		\item  Normalize $\textbf{w}_h$ to 1
		\item  $\textbf{t}_h = \textbf{X}_{h-1}\textbf{w}_h / \textbf{w}_h^t\textbf{w}_h$
		\item  $\textbf{q}_h = \textbf{X}_{h-1}^t\textbf{t}_h / \textbf{t}_h^t\textbf{t}_h$
		\item  $\textbf{X}_h = \textbf{X}_{h-1} - \textbf{t}_h\textbf{q}_h^t$
		\item  ${c}_h = \textbf{y}_{h-1}^t\textbf{t}_h / \textbf{t}_h^t\textbf{t}_h$
		\item	$\textbf{y}_h = \textbf{y}_{h-1} - {c}_h\textbf{t}_h^t$
	\end{enumerate}
\end{enumerate}

NIPALS-PLSR can therefore be seen as a compromise between a multiple linear regression and a principal component analysis, in which the  first $h$ components $\textbf{t}_h$ are the principal components whose covariances with $\textbf{y}$ are the largest. 

\subsection{Component selection and cross-validation with missing values}

Several papers have studied methods to determine the number $h$ of components to retain in the final PLS model---see for instance \cite{Lazraq2003}. In the present study, only selection of the number of components is considered, including all $\mathbf{x}_j$ variables on each of the first $h$ components, whatever their significance for the component. Several approaches in the literature  choose the number of components $h$ to include in a model, e.g., the $Q^2$ criterion, computed by cross-validation \citep{Stone1974}, the Akaike Information Criterion ($AIC$) \citep{Akaike1969}, and the Bayesian Information Criterion ($BIC$) \citep{Schwartz1978}, all of which we briefly describe below.

Using the NIPALS algorithm, it is possible to perform PLS regression even on incomplete data sets. When applied to incomplete data sets, cross-validation, either by leave-one-out or $k$-fold cross-validation, requires modifications. There are two ways to do this: \textit{standard} or \textit{adaptative} cross-validation. The first way is to predict the response value of any row of the dataset as if it contained missing data; here the cross-validation is said to be \textit{standard}. In \textit{adaptative} cross-validation, we predict the response value for a row accordingly to the presence (or not) of missing data in that row: regular prediction as for complete data if there are no missing data in the row, and missing data-specific prediction if there are missing data in the row \citep{Bertrand2015}.

The number of degrees of freedom (DoF) has been computed, for complete data, using the methods of \citet{Kramer2011} and implemented in the R package \texttt{plsdof} \citep{Kraemer2015}. The PLS routines in the  \texttt{plsRglm} package are based on these DoF except in the case of incomplete data for which only naive DoF are currently known.

The different criteria selected for our study are defined as follows:
\begin{itemize}
	\item $Q^2$. This statistic  is defined for each step $h$ as:
	\begin{equation}
		Q^2_{h} = 1-PRESS_h/RSS_{h-1}.
	\end{equation}
	For $h$ = 1, ${ RSS }_{ 0 }=\sum _{ i=1 }^{ n }{ { ({ y }_{ i }-{ \bar { y }  }) }^{ 2 } } =n-1$, where $\bar { y }$ is the mean of $\textbf{y}$ and ${ RSS }_{ h-1 }$ is the \emph{residual sum of squares} when the number of components is equal to $h-1$. $PRESS_h$ is the (\emph{predictive residual error sum of squares}) when the number of components is equal to $h$. A new component $\textbf{t}_h$ is considered significant for the prediction of $\textbf{y}$ if \citep[see][]{Tenenhaus1998}:
		\begin{equation}
			\sqrt { { PRESS }_{ h } } \le 0.95\sqrt { { RSS }_{ h-1 } } \Leftrightarrow  Q^2_h \ge 0.0975.
		\end{equation}	
	
	\item $AIC$ \citep{Akaike1974}. This is defined as:
	\begin{equation}
		AIC(m) = n log({ \hat { \sigma  }  }^{ 2 })+2m ,
	\end{equation}
	where $m$ is the number of model parameters, $n$ the sample size, and ${ \hat { \sigma  }  }^{ 2 }$ the maximum likelihood estimate of the variance of the response variable. 
	
	\item $BIC$ \citep{Schwartz1978}. This is defined as:
	\begin{equation}
		BIC=RSS/n+log(n)(\gamma /n){ \hat { \sigma  }  }_{ \epsilon  }^{ 2 },
	\end{equation}
	
	where $\gamma$ represents the $DoF$ of the model and ${ \hat { \sigma  }  }_{ \epsilon  }^{ 2 }$ the maximum likelihood estimate of the variance of the error.
	\cite{Kramer2011} defined $\gamma$ as the $DoF$ of the PLS regression with $h$ components and provided an unbiased estimate: 
	\begin{equation}
		{ \hat { DoF }  }{(h) }=1+\sum _{ j=1 }^{ h }{ { c }_{ j } \textup{trace}({\textbf{D}}^{ j }) }-\sum _{ { j,l=1 } }^{ h } { \textbf{t}}_{ { l } }^{ t }{ \textbf{D} }^{ { j } }{ \textbf{t}}_{ { l } }+({ \textbf{y} }-{ \hat { \textbf{y} }  }_{ h })^{ t }\sum _{j=1 }^{ h }{ { \textbf{D} }^{ j }{ \textbf{v} }_{ j } } + h,
	\end{equation}
	
	with $\textbf{D}=\textbf{X}{\textbf{X}}^{ t }$, $\textbf{B}=(< {\textbf{t}}_{ { h }_{ 1 } },{ \textbf{D} }^{ { h }_{ 2 } }\textbf{y} >)_{1\leqslant h_1,h_2 \leqslant h}$, $ \textbf{V}=({\textbf{v} }_{ 1 },\ldots,{ \textbf{v}}_{ h })=\textbf{T}{ ({ \textbf{B} }^{ -1 }) }^{ t }$ and $\textbf{c}={ \textbf{B}}^{ -1 }{\textbf{T}}^{ t }\textbf{y}.$
\end{itemize}

\subsection{Imputation methods} 
\subsubsection{Multiple imputation} 

Multiple imputation is a general statistical method for the analysis of incomplete datasets \mbox{\citep{Rubin1987,Royston2004,VanBuuren2012}.} This method has become a popular approach for dealing missing data in numerous analyses from different domains. The aim of multiple imputation is to provide unbiased and valid estimates of associations based on information from the available data.

The idea underlying multiple imputation is to use the observed data distribution to generate plausible values for the missing data, replacing them several times over several runs, then combining the results. The multiple imputation algorithm has three steps \citep{Rubin1996}. The first involves specifying and generating plausible values for missing values in the data. This stage, called imputation, creates multiple imputed datasets ($m$ of them). In the second step, a statistical analysis is performed on each of the $m$ imputed data set to estimate  quantities of interest. The results of the $m$ analyses will differ because the $m$ imputations differ. There is variability both within and between the imputed data sets because of the uncertainty related to missing values. The third step pools the $m$ estimates into one, combining both within- and between- imputation variation.

The question of the optimal number of imputations has been addressed by several authors. Rubin recommended 2 to 5 imputations in \citep{Rubin1987}. His argument was that even with 50\% missing data, five imputed data sets would produce point estimates that were 91\% as efficient as those based on an infinite number of imputations. In 1998, \cite{Graham2007} suggested 20 or more imputations. Later \citet{Bodner2008} and \cite{White2011} suggested the rule of thumb that $m$, the number of imputations, should be at least equal to the percentage of missing entries, which is what we do in this paper. 

Multiple imputation by chained equations (MICE) is a practical approach to generating imputation in the first step of multiple imputation \citep{VanBuuren2011}. A more detailed description of the theory involved is provided by \cite{VanBuuren2007}, \cite{VanBuuren2011}, and \cite{Azur2012}. In this study, we used the R package \texttt{mice} \citep{VanBuuren2018}.

\subsubsection{$k$-nearest neighbors imputation} 

The method of $k$-nearest neighbors imputation (KNNimpute) estimates a missing data point using values calculated from its $k$ nearest neighbors, defined in terms of similarity \citep{Dixon1979}. In particular, nearest neighbors can be with respect to some distance function. 

Types of distances that can be used include the Pearson correlation, Euclidean, Mahalanobis, Chebyshev, and Gower distances. Essentially, two far apart vectors are less likely than close together ones to have similar values. For a given missing data point, KNNimpute searches the whole dataset for its nearest neighbors. The missing value is then replaced by averaging the (non-missing) values of these neighbors. The  method's accuracy depends on the number of neighbors taken into account. Here we choose to use the Gower distance \citep[see][]{Kowarik2016}, which is coded in the R package \texttt{VIM} \citep{Templ2017}.

\subsubsection{Singular value decomposition imputation} 
The singular value decomposition imputation (SVDimpute) algorithm was proposed by \cite{Troyanskaya2001}. This algorithm estimates missing values as linear combinations of the $k$ most significant eigenvectors, where the most significant eigenvector is the one with the largest (in absolute value) eigenvalue. In this study, we used the R package \texttt{bcv} \citep{Perry2015} to run this.

\section{Simulation study}
\label{sec:2}
\subsection{Reference dataset construction}
Complete data sets with a defined number of components were generated using the method described in \cite{Li2002}. The true number of components was chosen to be 2, 4 or 6. The univariate response $\textbf{y}$ was distributed according to a Gaussian distribution $\mathcal{N}(0,1)$. Simulations were performed by adapting the \texttt{simul\_data\_UniYX} function available in the R package \texttt{plsRglm}. 

\subsection{Data dimensions}
PLS regression is particularly pertinent for data matrices in which $n < p$, but the behavior of the NIPALS algorithm can depend on whether $n < p$, or vice versa. Its properties have thus been studied on vertical data matrices, i.e., those for which $n > p$ (e.g., $n$ = 100 and $p$ = 20 in our study) and horizontal data matrices, i.e., those for which $n < p$ (e.g., $n$ = 20 and $p$ = 100 here). The range of scenarios we consider is shown in Section~\ref{sec:21}.

\subsection{Missing data and missingness mechanism}
Missing data were generated under MCAR  and MAR. The percentage of missing data $d$ took values in $d$ $\in \{5,10,\ldots,50\}$\%. Smaller proportions than 5\% of missing data could have been used for $n = 100$, but preliminary runs showed that the results were very close to those for $d = 5$\%.  Moreover, it was decided not to include missing data rates larger than 50\% in our study since it is more than questionable to run a model on a dataset in which more than half of the data are missing. 

\subsection{Simulation study design}
\label{sec:21}
\begin{figure}[t!]
	\centering
	\includegraphics[width=0.7\textwidth]{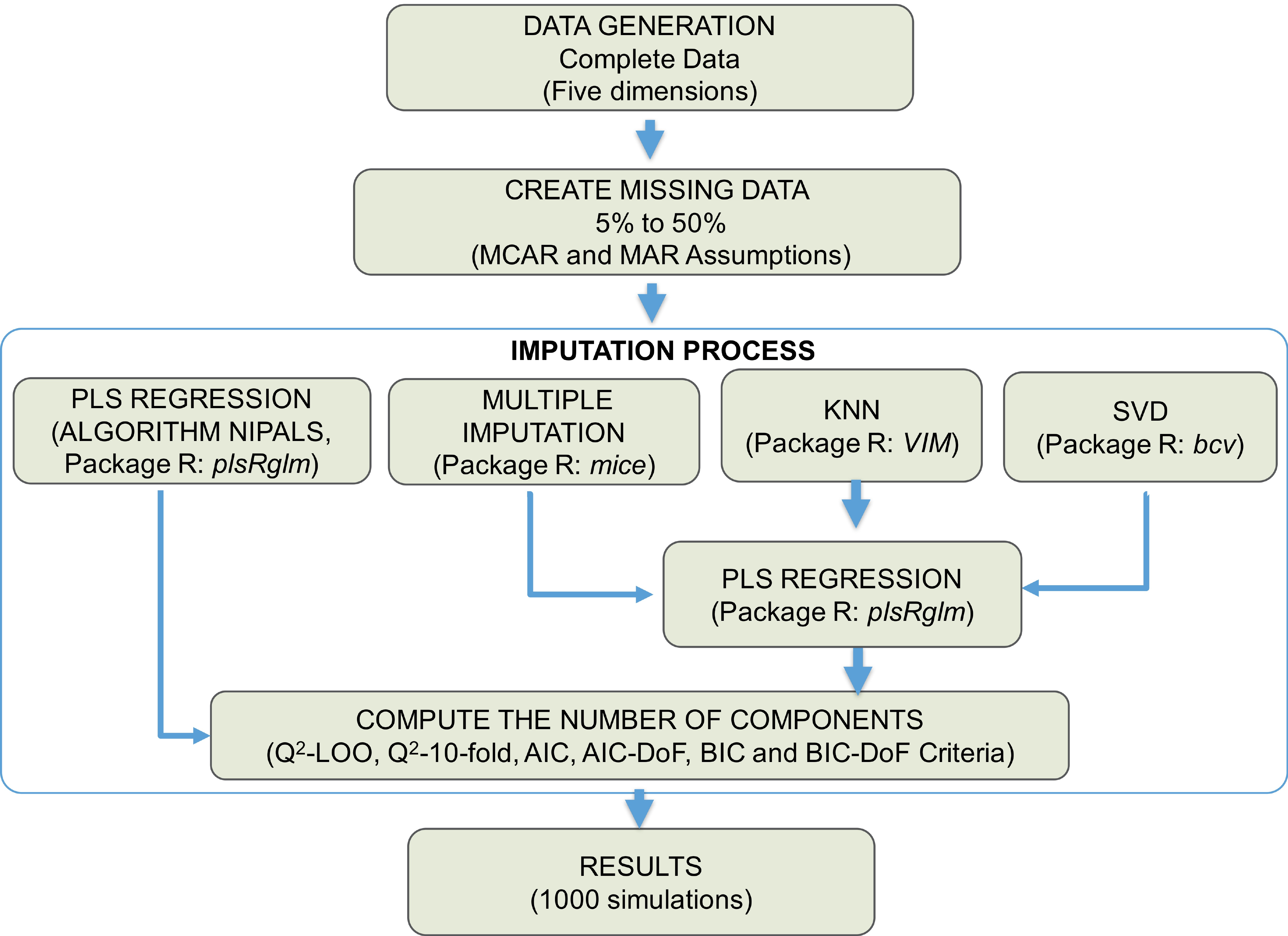}
	\caption{Simulation design.}
	\label{fig: simulation}
\end{figure}

The simulation study was designed as follows:
\begin{enumerate}
	\item Data were simulated as in \cite{Li2002}, with $n$ (number of observations), $p$ (number of variables), and the true number of components set to 2, 4 or 6 with each of the following five set-ups: 
	\begin{itemize}
		\item $n = 100$ and $p = 20$,
		\item $n = 80$ and $p = 25$,
		\item $n = 60$ and $p = 33$,
		\item $n = 40$ and $p = 50$,
		\item $n = 20$ and $p = 100$.
	\end{itemize}    
	\item Missing data were created under the MCAR and MAR assumptions, with the proportion of missing data going from 5\% to 50\% in steps of 5\%.     
	\item Missing data were imputed using MICE from the \texttt{mice} package and the \texttt{norm} imputation method, KNNImpute with the \texttt{VIM} package, and SVDimpute with the \texttt{bcv} package.      
	\item The number of components was computed using $Q^2$ leave-one-out cross-va\-li\-da\-tion and $Q^2$ 10-fold cross-validation computed on the incomplete data according to the \textit{standard} or \textit{adaptative} methods in the \texttt{PlsRglm} package. In the multiple imputation, the number of components was calculated---by $Q^2$ cross-validation---as the mode of the computed number of components across all $m$ imputations, where $m$ was equal to the percentage of missing data \citep{White2011}.
	\item We also computed the maximum number of components that could be extracted for PLS regression---which was up to 8 components---on incomplete data. Remember that the true number of components was either 2, 4 or 6.
\end{enumerate}
For each combination of (i) proportion of missing data, (ii) matrix dimensions, (iii) type of missingness, and (iv) true number of components, 1000 replicate datasets were drawn.

\section{Results}
\label{sec:3}
\begin{sidewaysfigure}[b!]
	\centering
	\includegraphics[width=\textwidth]{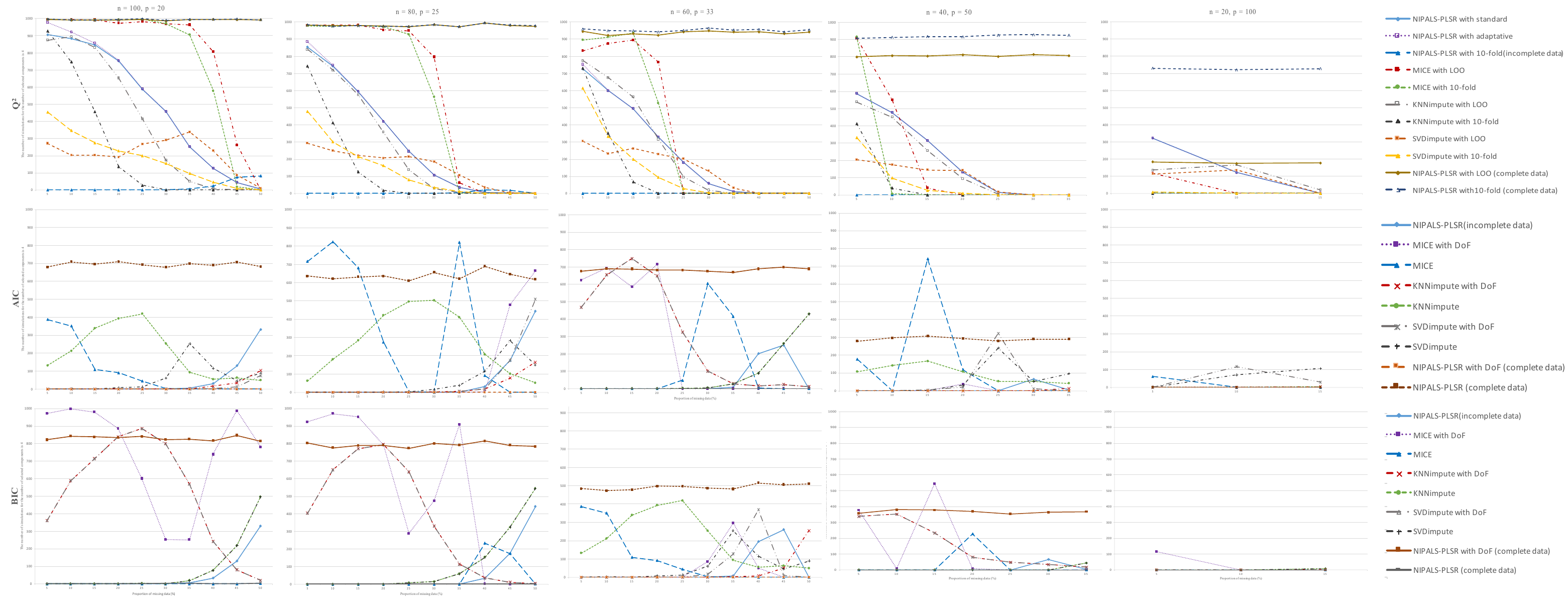}
	\caption{Evaluation of $Q^2$, $AIC$, and $BIC$ for NIPALS-PLSR, MICE, KNNimpute, and SVDimpute under MCAR. Results are expressed as the number of simulations for which the selected number of components was 4 (the true value).}
	\label{fig: MCAR}
\end{sidewaysfigure}

\begin{sidewaysfigure}[t!]
	\centering
	\includegraphics[width=\textwidth]{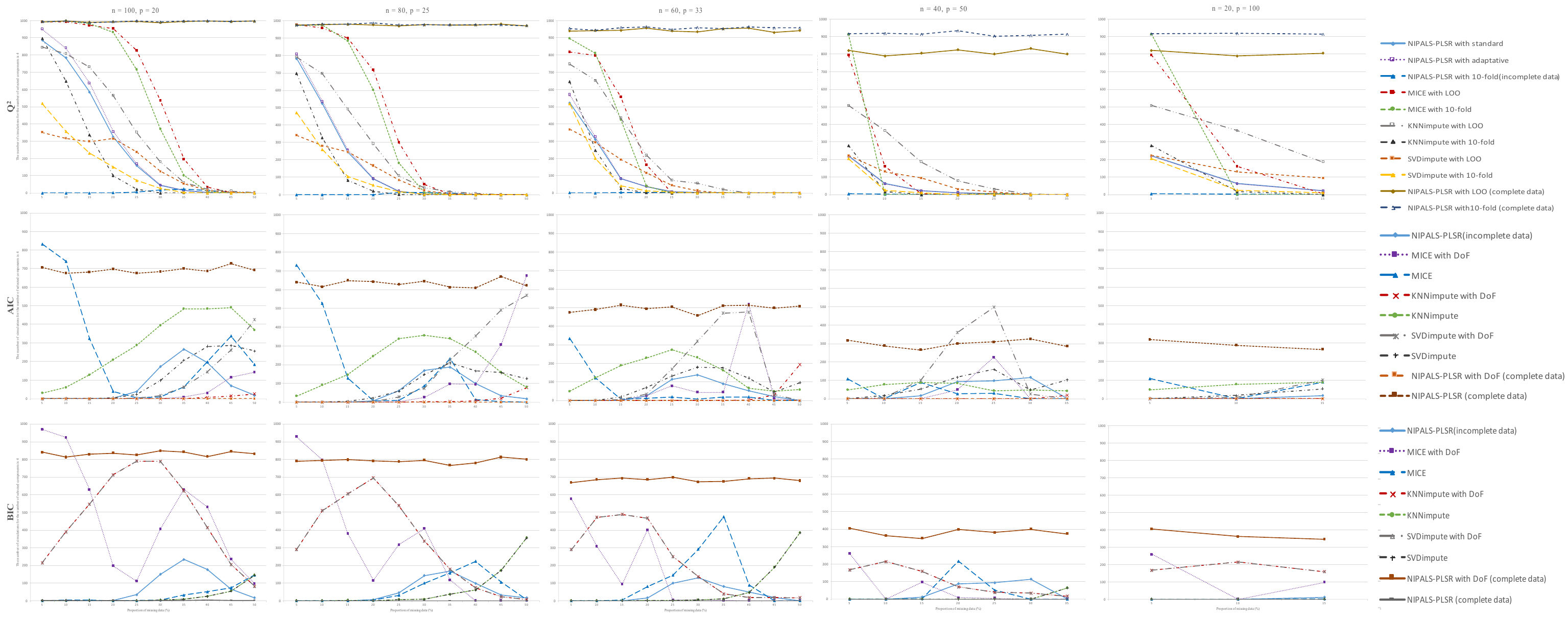}
	\caption{Evaluation of $Q^2$, $AIC$, and $BIC$ for NIPALS-PLSR, MICE, KNNimpute, and SVDimpute under MAR. Results are expressed as the number of simulations for which the selected number of components was 4 (the true value).}
	\label{fig: MAR}
\end{sidewaysfigure}

Figures \ref{fig: MCAR} and \ref{fig: MAR} plot the performance of each method as a function of the proportion of missing data with various shaped matrices under both MCAR and MAR.  These results show that the performance of $Q^2$-LOO in selecting the correct number of components increases as the  sample size does, and as expected, decreases as the proportion of missing data increases for both MCAR and MAR. MICE with $Q^2$-LOO gave results closest to the correct number of components on the incomplete data set under both the MCAR and MAR assumptions, followed by NIPALS-PLSR and KNNimpute, when the proportion of missing data was small ($<$ 30\%), in the vertical matrix setting. SVDimpute with $Q^2$-LOO performed the worst: the true number of components was correctly selected in only around one third of the simulations.

We see that the $Q^2$-10-fold performed less well, and overall, selected the correct number of components less frequently than $Q^2$ did. In the vast majority of situations (i.e.,  combinations of matrix size, proportion, and pattern of missing data), the number of components it selected was on average greater than the true number.  

We also found that $AIC$, $AIC$-DoF, and $BIC$ systematically selected a larger number of components than $Q^2$. This difference can sometimes be as large as three or four for each of the methods and  both the MCAR and MAR cases. 

The actual number of  components selected by either MICE, KNNimpute, or SVDimpute with the $BIC$-DoF criterion was systematically larger than the number of components selected by $BIC$, $AIC$, and $AIC$-DoF. $BIC$-DoF's performance increases and then decreases as the  proportion of missing data increases, instead of decreasing regularly over the whole range of missing data proportions.

\section{Discussion and conclusion}
\label{sec:4}
PLS regression is a multivariate method for which two algorithms (SIMPLS and NIPALS) can be used to provide the model's parameter estimates. The NIPALS algorithm has the interesting property of being able to provide estimates on incomplete data; this has been extensively studied in the case of PCA---for which  NIPALS was originally devised.

Here, we have studied the behavior of the NIPALS algorithm when used to fit PLS regression models for various proportions of missing data and for different types of missingness. Comparisons with MICE, KNNimpute, and SVDimpute were performed under the MCAR and MAR assumptions. In our simulations, the model dimension (i.e., the optimal number of components) was computed according to different criteria, including $Q^2$, $AIC$, and $BIC$. 

The number of selected components, be it for complete, incomplete, or imputed datasets, depends on the criterion used. The fact that $AIC$ and $BIC$ select a larger number of components than $Q^2$, has been already observed in another setting by the present authors \citep{Meyer2010} and by others \citep{Li2002}.

The number of components selected using MICE is much closer to the correct number of components when the proportion of missing data is small ($<$ 30\%) and for vertical matrices. However, the MICE computation time was long, and depended on the proportion of missing data, increasing as the proportion of missing data did. For instance, the average MICE run time for $n = 100 \times p= 20$, was about 11 times longer than that of NIPALS-PLSR when $d = 10\%$, and around 40 times longer when $d = 50\%$ under MCAR. In contrast, NIPALS-PLSR, KNNimpute, and SVDimpute, had very fast run times:  0.5--1.5 seconds on average. Generally, the run time under MAR was longer than under MCAR for both vertical and horizontal matrices. Consequently, though MICE may be the method of choice, its run time may prohibit its use in practice.

$BIC$-DoF, a criterion derived from $BIC$, gives a slightly better estimation of the number of components when the proportion of missing data is small, particularly for MICE. This shows that taking into account a modified number of DoF can substantially improve the likelihood of selecting the correct number of components \citep{Kramer2011}. Further research would nevertheless be useful to extend this version of $BIC$ to other settings like for instance GLM, or adapt it to specific cases of incomplete datasets that require further $DoF$ adjustments.

For smaller sample sizes $n$, the multivariate structure of the data was not taken into account in the imputations due to high levels of collinearity. Indeed, the smaller the sample size, the more difficult it was for the MICE algorithm to converge. Thus, though it would have been possible to run the imputation, the PLS regression estimates would have been biased. This implies that our conclusions for very small sample sizes may be misleading. Such biased parameter estimates could also bias the comparisons between the methods, but also hint at the fact that even a small proportion of missing data can make it difficult to estimate the correct number of components in PLS regression.

\begin{table}[t!]
	\centering
	\caption{The results of our study based on seven evaluation criteria when the true number of components is set to 2, 4, or 6 under MCAR and vertical matrices. The number of "+" indicates performance, from poor (+) to very good (++++).}
	\label{table:1}  
	\begin{adjustbox}{width=\textwidth}
		\begin{tabular}{ccccccccccccc}
			\hline\noalign{\smallskip}	
			&\multicolumn{4}{c}{Two-Components}	& \multicolumn{4}{c}{Four-Components}& \multicolumn{4}{c}{Six-Components}\\		
			\cmidrule(lr){2-5} \cmidrule(lr){6-9} \cmidrule(lr){10-13}
			Criteria & NIPALS-PLSR &MICE	&KNNimpute	&SVDimpute	& NIPALS-PLSR &MICE	&KNNimpute	&SVDimpute & NIPALS-PLSR &MICE	&KNNimpute	&SVDimpute\\
			\noalign{\smallskip}\hline\noalign{\smallskip}		
			$Q^2$-LOO&++++&+++&++++&+++&++&+++&++&+&+&++&+&+\\
			$Q^2$-10-fold&+&+++&++++&++++&+&+++&++&+&+&++&+&+\\
			$AIC$&+&+&+&+&+&+&+&+&+&+&+&+\\
			$AIC$-DoF&-&++&+&+&-&++&+&+&-&+&+&+\\
			$BIC$&+&+&+&+&+&+&+&+&+&+&+&+\\
			$BIC$-DoF&-&+++&+++&++&-&+++&+++&+&-&++&+&+\\
			Execution time &++++&+&++&+++&++++&+&++&+++&++++&+&++&+++\\
			\noalign{\smallskip}\hline
		\end{tabular}
	\end{adjustbox}
\end{table}

\begin{table}
	\centering
	\caption{The results of our study based on seven evaluation criteria when the true number of components is set to 2, 4, or 6 under MCAR and horizontal matrices. The number of "+" indicates performance, from poor (+) to very good (++++).}
	\label{table:2}  
	\begin{adjustbox}{width=\textwidth}
		\begin{tabular}{ccccccccccccc}
			\hline\noalign{\smallskip}	
			&\multicolumn{4}{c}{Two-Components}	& \multicolumn{4}{c}{Four-Components}& \multicolumn{4}{c}{Six-Components}\\			
			\cmidrule(lr){2-5} \cmidrule(lr){6-9} \cmidrule(lr){10-13}
			Criteria & NIPALS-PLSR &MICE	&KNNimpute	&SVDimpute	& NIPALS-PLSR &MICE	&KNNimpute	&SVDimpute & NIPALS-PLSR &MICE	&KNNimpute	&SVDimpute\\
			\noalign{\smallskip}\hline\noalign{\smallskip}		
			$Q^2$-LOO&+++&++&++&++&++&++&++&+&+&+&+&+\\
			$Q^2$-10-fold&+&++&+++&+++&+&++&+&+&+&+&+&+\\
			$AIC$&+&+&+&+&+&+&+&+&+&+&+&+\\
			$AIC$-DoF&-&+&+&+&-&+&+&+&-&+&+&+\\
			$BIC$&+&+&+&+&+&+&+&+&+&+&+&+\\
			$BIC$-DoF&-&++&+&+&-&+&+&+&-&+&+&+\\
			Execution time &+++&+&++&++++&+++&+&++&++++&+++&+&++&++++\\
			\noalign{\smallskip}\hline
		\end{tabular}
	\end{adjustbox}
\end{table}

\begin{table}
	\centering
	\caption{The results of our study based on seven evaluation criteria when the true number of components is set to 2, 4, or 6 under MAR and vertical matrices. The number of "+" indicates performance, from poor (+) to very good (++++).}
	\label{table:3}  
	\begin{adjustbox}{width=\textwidth}
		\begin{tabular}{ccccccccccccc}
			\hline\noalign{\smallskip}	
			&\multicolumn{4}{c}{Two-Components}	& \multicolumn{4}{c}{Four-Components}& \multicolumn{4}{c}{Six-Components}\\			
			\cmidrule(lr){2-5} \cmidrule(lr){6-9} \cmidrule(lr){10-13}
			Criteria & NIPALS-PLSR &MICE	&KNNimpute	&SVDimpute	& NIPALS-PLSR &MICE	&KNNimpute	&SVDimpute & NIPALS-PLSR &MICE	&KNNimpute	&SVDimpute\\
			\noalign{\smallskip}\hline\noalign{\smallskip}		
			$Q^2$-LOO&+++&+++&++++&++++&++&+++&++&+&+&++&+&+\\
			$Q^2$-10-fold&+&+++&++++&+++&+&+++&++&++&+&++&+&+\\
			$AIC$&+&+&+&+&+&+&+&+&+&+&++&+\\
			$AIC$-DoF&-&++&++&+&-&++&+&+&-&+&+&+\\
			$BIC$&+&+&+&+&+&+&+&+&+&+&++&+\\
			$BIC$-DoF&-&+++&+++&+&-&+++&+++&+&-&+&+&+\\
			Execution time &++++&+&+++&++++&++++&+&+++&++++&++++&+&+++&++++\\
			\noalign{\smallskip}\hline
		\end{tabular}
	\end{adjustbox}
\end{table}

\begin{table}
	\centering
	\caption{The results of our study based on seven evaluation criteria when the true number of components is set to 2, 4, or 6 under MAR and horizontal  matrices. The number of "+" indicates performance, from poor (+) to very good (++++).}
	\label{table:4}  
	\begin{adjustbox}{width=\textwidth}
		\begin{tabular}{ccccccccccccc}
			\hline\noalign{\smallskip}	
			&\multicolumn{4}{c}{Two-Components}	& \multicolumn{4}{c}{Four-Components}& \multicolumn{4}{c}{Six-Components}\\			
			\cmidrule(lr){2-5} \cmidrule(lr){6-9} \cmidrule(lr){10-13}
			Criteria & NIPALS-PLSR &MICE	&KNNimpute	&SVDimpute	& NIPALS-PLSR &MICE	&KNNimpute	&SVDimpute & NIPALS-PLSR &MICE	&KNNimpute	&SVDimpute\\
			\noalign{\smallskip}\hline\noalign{\smallskip}		
			$Q^2$-LOO&+&++&+++&+++&+&++&++&+&+&+&+&+\\
			$Q^2$-10-fold&+&++&+++&+++&+&+&+&+&+&+&+&+\\
			$AIC$&+&+&+&+&+&+&+&+&+&+&+&+\\
			$AIC$-DoF&-&+&+&+&-&+&+&+&-&+&+&+\\
			$BIC$&+&+&+&+&+&+&+&+&+&+&+&+\\
			$BIC$-DoF&-&+&+&+&-&+&+&+&-&+&+&+\\
			Execution time &++++&+&+++&++&++++&+&+++&++&++++&+&+++&++\\
			\noalign{\smallskip}\hline
		\end{tabular}
	\end{adjustbox}
\end{table}

The framework for our simulations was based on those in \citet{Li2002}, with the true number of components set to either 2, 4 or 6. A summary of these results for all criteria are shown in Tables~\ref{table:1}--\ref{table:4}. The first row gives the performance of $Q^2$-LOO when the true number of components is 2, then 4, and finally 6. In this case, the $Q^2$-LOO criterion is more efficient than for the four-component and six-component models. For the two-component case, NIPALS-PLSR and KNNimpute with $Q^2$-LOO provide a satisfactory performance under MCAR, whereas under MAR, KNNimpute and SVDimpute with $Q^2$-LOO are the most efficient. It is thus clear that the $Q^2$-LOO criterion has the best performance. Theoretically, the leave-one-out method can extract the maximum possible information \citep{Eastment1982}. This result is supported by the simulation results in this study.

In conclusion, our simulations show that whatever the criterion used, the type of missingness and proportion of missing data must also be taken into consideration since they both influence the number of components selected. The true number of components of a PLS regression was difficult to determine, especially for small sample sizes and when the proportion of missing data was larger than 30\%. Moreover, under MCAR, the true number of selected components using these methods was generally closer to the true number of components than in the MAR setting. 

\clearpage

\bibliographystyle{apalike}
\bibliography{ref}

\end{document}